\documentclass[aps,pre,twocolumn,showpacs]{revtex4}
\usepackage{epsfig}
\usepackage{times}

\begin{document}

\title{Effects of competition on pattern formation in the rock-paper-scissors game}

\author{Luo-Luo Jiang,$^{1,2}$ Tao Zhou,$^{3}$ Matja{\v z} Perc,$^4$ Bing-Hong Wang$^{5}$}
\affiliation{$^1$College of Physics and Electronic Information Engineering, Wenzhou
University, Wenzhou 325027, China\\
$^2$College of Physics and Technology, Guangxi Normal University, Guilin
541004, China\\
$^3$Web Sciences Center, University of Electronic Science and Technology of
China, Chengdu 610051, China\\
$^4$Faculty of Natural Sciences and Mathematics, University of Maribor,
Koro{\v s}ka cesta 160, SI-2000 Maribor, Slovenia\\
$^5$Department of Modern Physics, University of Science and Technology of
China, Hefei 230026, China}

\begin{abstract}
We investigate the impact of cyclic competition on pattern formation in the rock-paper-scissors game. By separately considering random and prepared initial conditions, we observe a critical influence of the competition rate $p$ on the stability of spiral waves and on the emergence of biodiversity. In particular, while increasing values of $p$ promote biodiversity, they may act detrimental on spatial pattern formation. For random initial conditions, we observe a phase transition from biodiversity to an absorbing phase, whereby the critical value of mobility grows linearly with increasing values of $p$ on a log-log scale, but then saturates as $p$ becomes large. For prepared initial conditions, we observe the formation of single-armed spirals, but only for values of $p$ that are below a critical value. Once above, the spirals break up and form disordered spatial structures, mainly because of the percolation of vacant sites. Thus, there exists a critical value of the competition rate $p_{c}$ for stable single-armed spirals in finite populations. Importantly though, $p_{c}$ increases with increasing system size, because noise reinforces the disintegration of ordered patterns. In addition, we also find that $p_{c}$ increases with the mobility. These phenomena are reproduced by a deterministic model that is based on nonlinear partial differential equations. Our findings indicate that competition is vital for the sustenance of biodiversity and emergence of pattern formation in ecosystems governed by cyclical interactions.
\end{abstract}

\maketitle

\section{Introduction}

Biodiversity and spatiotemporal dynamics of interacting individuals are important for the characterization of ecological systems. Recently, spatial heterogeneity of species has attracted much attention because it is closely related to the stability and coexistence in ecological and evolutionary systems \cite{hofbauer_98, nowak_06, sigmund_10, szabo_pr07, roca_plr09, perc_bs10}, whereas the competition between species was for example invoked as an evolutionary force in community food webs \cite{stanley_pnas73, vermeij_87}. Since each species has a different competition rate with which it is able to invade others, Huntley and Kowaleski studied Phanerozoic fossil records and found that the competition rate is deeply interrelated with biodiversity \cite{huntley_pnas07}. Theoretically, several aspects of multi-species cyclical dominance have already been studied in detail. For example, it has been established that three species in a cyclic dominance exhibit self-organizing behavior on the spatial grid \cite{tainaka_prl89, frachebourg_pre96}, whereby similar observations can be made also for system that incorporate more than three species, provided their total number does not exceed fourteen \cite{frachebourg_jpa98}. Phase transitions and selection have also been studied in the predator-prey models allowing some motion throughout inhabitable vacant sites \cite{szabo_pre04, he_m_ijmpc05}. Recent theoretical studies focused on biodiversity and the emergence of self-organizing patterns by competing associations with homogeneous and heterogeneous invasion rates \cite{malcai_pre02, provata_pre03, lai_prl05, szabo_pre08, reichenbach_n07, szabo_pre08b, perc_njp07b, reichenbach_prl07, perc_pre07b, claussen_prl08, jiang_ll_njp09, berr_prl09, szolnoki_pre10b, yang_r_c10, wang_wx_pre10b, ni_x_pre10, shi_hj_pre10, case_epl10, winkler_pre10, wenxu_pre11}, elevating models of cyclical interactions to a very vibrant and enthusiastically approached field of research.

Despite being fascinatingly simple and seemingly trivial, cyclical interactions among competing species emerge rather frequently in nature, and are indeed becoming more and more established as excellent models for the study of biodiversity, the formation of defensive alliances \cite{szabo_pre01b} and Darwinian selection \cite{maynard_n73}, as well as structural complexity \cite{watt_je47} and prebiotic evolution in general \cite{rasmussen_s04}. Examples of reported real-life occurrences of cyclical interactions include rodents in the hight-Arctic tundra in Greenland \cite{gilg_s03}, lizards in the inner Coast Range of California \cite{sinervo_n96}, overgrowths by marine sessile
organisms \cite{burrows_mep98}, and microbial populations of $\textit{Escherichia coli}$ \cite{kirkup_n04, kerr_n02}. Recent experiments revealed that cyclical interactions can promote biodiversity within three strains of $\textit{Escherichia coli}$ \cite{kerr_n02}. However, while effects of mobility and noise on cooperation, biodiversity and spatial pattern formation have been investigated intensely and with great success \cite{vainstein_pre01, reichenbach_n07, perc_njp06a, vainstein_jtb07, helbing_pnas09, meloni_pre09, droz_epjb10, venkat_pre10}, the effects of different rates of cyclic competition have not yet been thoroughly investigated. It is thus of interest to systemically investigate the impact of cyclic competition rate on both, biodiversity and pattern formation in models of cyclically competing species.

Accordingly, we investigate effects of different competition rates on phase transitions and pattern formation by employing both random and prepared initial conditions. With increasing of mobility, the phase transition from biodiversity to an absorbing phase, where two species go extinct, emerges, and the critical value of mobility is critically affected by the competition rate. Large competition rates lead to vacant sites surrounding small patches containing all three species, which results in the fragmentation of macroscopical spirals. This effect prevents the pattern outgrowing the system size, and in turn inducing promotion of biodiversity. For random initial conditions our results are in agrement with the seminal results of Reichenbach \textit{et al.} \cite{reichenbach_n07, reichenbach_prl07}. For prepared initial conditions, however, we find that with increasing of the competition effects, a sharp transition occurs from a single spiral to multiple spirals coexisting. The disintegration is due to the percolation of vacant sites as well as noise. Since the level of the latter is directly related to the system size, the critical competition rate $p_{c}$ increases with the increasing system size $L \times L$, such that it is convenient to define the so-called competition effect as $E=p/L$, of which the critical value $E_{c}$ (at which the single-arm spiral disintegrate) is then independent of $L$. For a fixed system size, however, we also find that the value of the critical competition increases with increasing mobility. It is thus demonstrated that the rate of cyclic competition may have positive effects on the promotion of biodiversity, but negative effects on the sustenance of single-armed spiral waves. The latter phenomenon is investigated in detail also for different reproduction rates governing the rock-paper-scissors game.

The remainder of this paper is organized as follows. In the next section we present the employed rock-paper-scissors game. Main results on effects of competition on biodiversity and pattern formation are presented in section 3. Section 4 features results on the impact of different reproduction rates on the emergence and stability of spirals. The last section summarizes the findings and discusses their potential implications.

\section{The rock-paper-scissors game}

We employ the biological rock-paper-scissors game with the following specifications. Nodes of a $L \times L$ square lattice present mobile individuals belonging to one of the three species, which we denote by $A$, $B$ and $C$. Each node can either host one individual of a given species or it can be vacant. Vacant sites, which we denote by $\otimes$, are also the so-called resource sites.
\begin{eqnarray}
AB \stackrel{p}{\longrightarrow} A\otimes&\,, \quad BC
\stackrel{p}{\longrightarrow} B\otimes&\,, \quad CA
\stackrel{p}{\longrightarrow} C\otimes\, \\
A\otimes \stackrel{q}{\longrightarrow} AA&\,, \quad B\otimes
\stackrel{q}{\longrightarrow} BB&\,, \quad C\otimes
\stackrel{q}{\longrightarrow}CC\, \\
A\odot \stackrel{\gamma}{\longrightarrow} \odot A&\,, \quad B\odot
\stackrel{\gamma}{\longrightarrow} \odot B&\,, \quad C\odot
\stackrel{\gamma}{\longrightarrow} \odot C\,
\end{eqnarray}
where $\odot$ denotes any species or vacant sites. These reactions
describe three processes, \textit{i.e.} competition, reproduction and
exchange, occurring only between neighboring nodes. In reaction (1), species $A$ eliminates species $B$ at a rate $p$, whereby the node previously hosting species $B$ becomes vacant. In the same manner species $B$ can eliminate species $C$, and species $C$ can eliminate species $A$, thus forming a closed loop of dominance between them. Reaction (2) shows that individuals can place an offspring to a neighboring vacant node $\otimes$ at a rate $q$. Reaction (3) defines an exchange process, whereat an individual may exchange its position with an individual belonging to any other species or an empty site at a rate $\gamma$.

Here we use both the stochastic and the deterministic approach to simulate the rock-paper-scissors game. For the stochastic approach, we make use of a stochastic simulation algorithm whereby the temporal evolution can be considered as a random walk process. The most commonly applied stochastic simulation algorithm was developed by Gillespie \cite{gillespie_jcp76, gillespie_jpc77}, where reactions occur in a random manner. In particular, competition occurs with probability $p /(p +q+\gamma)$, reproduction with probability $1 /(p +q+\gamma)$, and exchange (moving) with probability $\gamma /(p +q+\gamma)$. According to the random walk theory \cite{redner_01}, the mobility of individuals $M$ can be defined as $M=\gamma/2N$, meaning it is proportional to the typical area explored by a mobile individual per unit time.

As derived in the works of Reichenbach \textit{et al.} \cite{reichenbach_n07, reichenbach_prl07, reichenbach_prl08}, for the deterministic approach we use partial differential equations (PDE) of the form:
\begin{eqnarray}
\begin{array}{lll}
\partial_t a(\mathbf{r},t)=D\nabla^2a(\mathbf{r},t)+qa(\mathbf{r},t)\rho_{0}-pc(\mathbf{r},t)a(\mathbf{r},t),\\
\partial_t b(\mathbf{r},t)=D\nabla^2b(\mathbf{r},t)+qb(\mathbf{r},t)\rho_{0}-pa(\mathbf{r},t)b(\mathbf{r},t),\\
\partial_t c(\mathbf{r},t)=D\nabla^2c(\mathbf{r},t)+qc(\mathbf{r},t)\rho_{0}-pb(\mathbf{r},t)c(\mathbf{r},t),
\end{array}
\label{eq:PDE}
\end{eqnarray}
where $D$ denotes the diffusion rate and $\rho_{0}$ the density of vacant sites.

\section{Effects of competition on biodiversity and pattern formation}

\begin{figure}
\begin{center} \includegraphics[width = 8.5cm]{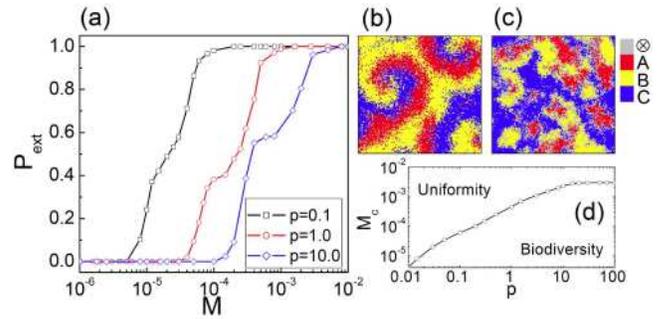}
\caption{\label{random}(color online) (a) Extinction probability $P_{ext}$ as a function of mobility $M$ for different competition rates $p$. As $M$ increases,
there is a transition from stable coexistence ($P_{ext}=0$) to extinction ($P_{ext}=1$). Panels (b) and (c) depict typical snapshots after a long relaxation period of the system for $p=1.0$ and $p=10.0$, respectively, at $M=1.0\times 10^{-4}$. (d) Phase diagram depicting the critical mobility $M_{c}$ as a function of the competition rate $p$, separating the absorbing single-species phase (uniformity) and the biodiversity phase. Random initial conditions, $N=128^{2}$, and $q=1.0$ were used for results in all panels.}
\end{center}
\end{figure}

To examine effects of the competition rate $p$ on the sustenance of biodiversity, we fix the reproduction rate $q$ equal to $1.0$ (see next section for the relaxation of this condition), and we perform extensive computer simulations of reactions (1)-(3) with no-flux boundary conditions, using first random initial configurations where each lattice site is occupied by an individual of species $A$, species $B$, species $C$, or is left empty with equal probability. Following Ref.~\cite{reichenbach_n07}, we calculate the extinction probability $P_{ext}$ that two species have gone extinct when the system reaches the stationary state. Figure~\ref{random}(a) features the results for $p=1.0$, which are in agrement to those reported in Ref. \cite{reichenbach_n07}, Namely, with increasing of $M$, a phase transition from biodiversity ($P_{ext}=0$) to uniformity ($P_{ext}=1$) emerges at a critical value $M_{c}=(4.5 \pm 0.5) \times 10^{-4}$. From Fig.~\ref{random}(a), one can also observe that the critical value of $M_{c}$ depends on $p$; in particular, smaller $p$ yield smaller values of $M_{c}$. To study how the competition rate $p$ affects the biodiversity, we present the typical snapshots for $p=1.0$ and $p=10.0$ in Figs.~\ref{random}(b) and (c), respectively, at $M=1.0\times 10^{-4}$. We find that while macroscopical spirals exist in the system at $p=1.0$, at $p=10.0$ small patches occupied by individuals of any of the three species are divided (or ``disconnected'') by vacant sites, and hence no macroscopical (large) spiral can be observed. According to Ref.~\cite{reichenbach_n07}, the loss of biodiversity results from spirals outgrowing the system size when mobility $M$ exceed $M_{c}$. Therefore, for a given $M$, $M=1.0\times 10^{-4}$ for example, the larger value of $p$ promotes biodiversity better than smaller $p$. In Fig.~\ref{random}(d), we present the phase diagram with the biodiversity phase and the absorbing single-species (uniformity) phase delineated. There are in fact two different regimes inferable, which are due to small and large values of $p$. For small values of $p$, \textit{i.e.} when the reproduction rate is comparable to the competition rate, $M_{c}$ grows linearly with $p$ on a log-log scale. However, for large values of $p$, \textit{i.e.} when the competition rate is significantly larger than the reproduction rate, the reproduction process limits the dynamics and therefore $M_{c}$ disobeys the linear relation and approaches a constant value for $p>10$.

\begin{figure}
\begin{center} \includegraphics[width = 8.5cm]{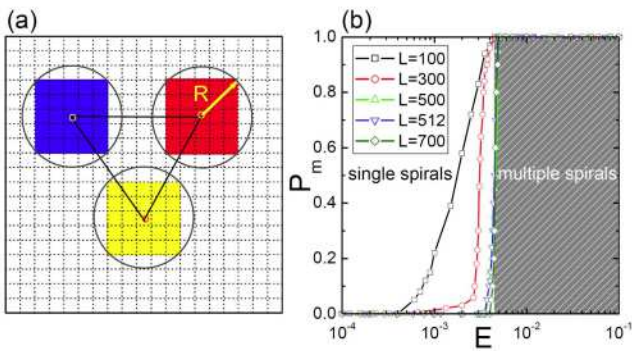}
\caption{\label{transitionheter}(color online) (a) Illustration of the realization of prepared initial conditions. Here $R = 3.5$, and every circle is occupied exclusively with a single species corresponding to the denoted color. All other sites are left vacant. Note that the distance between any two circles is the same. The radius $R$ is fixed to $10.5$ throughout this paper. (b) Phase transitions from single-arm spirals to coexisting multiple spirals, determined by means of $P_m$, in dependence on $E$ for $M=5.0\times 10^{-5}$, $q=1.0$ and different system size $N=L^2$. Here $E$ describes the so-called competition effect, which we introduce as $p/L$. Prepared initial conditions, as depicted in panel (a), were used.}
\end{center}
\end{figure}

Since pattern formation plays a central role in the sustenance of biodiversity \cite{reichenbach_n07,reichenbach_prl07}, we focus on the evolutionary process of pattern formation in dependence on the competition rate $p$. Thus, we will consider $p$ and the mobility $M$ as the two crucial parameters effecting pattern formation in the examined model. As shown in Fig.~\ref{transitionheter}(a), prepared initial condition that facilitate the emergence of spirals are employed for this purpose. In particular, three roundish areas with radius of $10.5$ are occupied individually by each of the three species, whereby the distance between each this region is the same. All other nodes are initially vacant, \textit{i.e.} providing resources needed for reproduction. Notably, such a setup has been considered previously in ecological systems, for example those aiming for experimental bacteria growth \cite{cyrill_pnas07}, and is frequently referred to as growth initial conditions setup.

\begin{figure}
\begin{center} \includegraphics[width = 8.5cm]{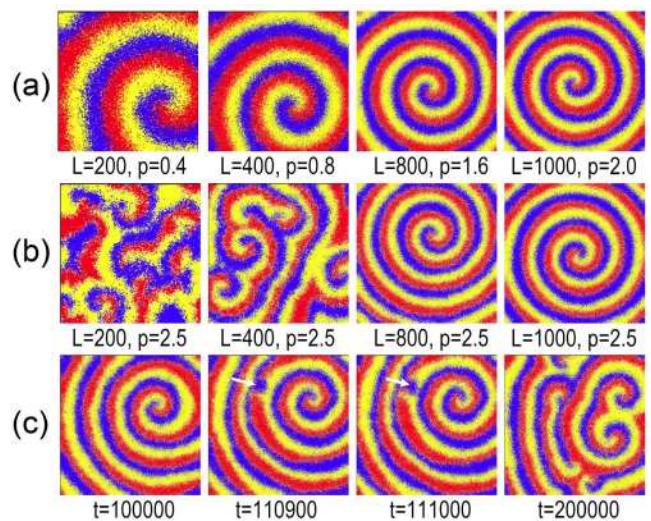}
\caption{\label{fragments}(color online) Panels in row (a) show single spirals at $E=2.0\times 10^{-3}$ for different system sizes. Panels in row (b) display spatial patterns at $p=2.5$ for $L=200$, $L=400$, $L=800$, and $L=1000$. Accordingly, for $L=200$ and $L=400$, because the competition effect $E$ is $1.25\times 10^{-2}$ and $6.25 \times 10^{-3}$, respectively (which exceeds $E_{c}$), multiple spirals are observed. However, for $L=800$ and $L=1000$, since the competition effect $E$ is $3.125\times 10^{-3}$ and $2.5 \times 10^{-3}$, respectively (which is below $E_{c}$), single spirals are stable. Temporal evolution of typical spatial patterns, as emerging for $p=2.5$ and $N=512^{2}$, is presented in row (c). At $t=100000$, a single-arm spiral emerges when the system is started with prepared initial conditions. Subsequently, the large spiral starts to break up at $t=110900$ and $t=111000$ (indicated by the white arrow). When reaching the stationary state at $t=200000$, the single-arm spiral observed at $t=100000$ is fragmented. Here $M=5.0\times 10^{-5}$ and $q=1.0$ in panels (a), (b) and (c).}
\end{center}
\end{figure}

\begin{figure}
\begin{center} \includegraphics[width = 8.0cm]{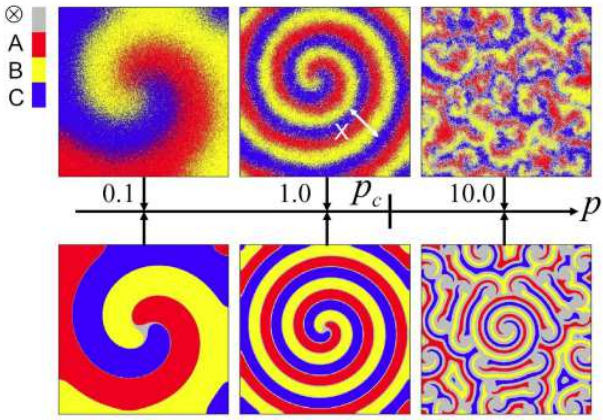}
\caption{\label{cri_predation}(color online) Typical spatial patterns emerging for different competition rates $p$ at $M=5.0 \times 10^{-5}$ (top panels) and $D=5.0 \times 10^{-5}$ (bottom panels). Prepared initial conditions, as depicted in Fig.~\ref{transitionheter}(a) were used. From left to right, the values of $p$ are $0.1$, $1.0$ and $10.0$, respectively. If $p< p_{c}=(2.3 \pm 0.3)$ single-arm spirals emerge, which are characterized by decreasing spatial wavelengths $\lambda=X/L$ as $p$ increases. Upon surpassing the critical value, the large spirals disintegrate into several fragmented spirals, forming essentially disordered spatial portraits. The upper three panels are obtained from Monte Carlo simulations, while the lower three panels depict results from the PDE model. Here $N=512^2$ and $q=1.0$.}
\end{center}
\end{figure}

We first investigate pattern formation by means of direct simulations of the game via the Monte Carlo method for different values of $p$. An elementary Monte Carlo step consists of randomly choosing an individual who interacts with one of its four nearest neighbors, which is also selected randomly, and then executing the process as determined by the Gillespie's algorithm \cite{gillespie_jcp76, gillespie_jpc77}. One full Mote Carlo step consists of $N=L^2$ elementary steps, during which, in accordance with the random sequential update, each player is selected once on average. Since noise increases with the decreasing of the system size, which on the other hand affects pattern formation, we define the so-called competition effect as $E=p/L$. In Fig.~\ref{transitionheter}(b), we plot the probability $P_m$ of multiple spirals coexisting as obtained in dependence on $E$ for $M=5.0\times 10^{-5}$. More accurately, $1-P_{m}$ is defined as the probability that the spatial grid is occupied by a single one-armed spiral. A sharp phase transition at a critical value $E_{c}=(4.5 \pm 0.5) \times 10^{-3}$ is observed. Because the transition becomes sharper with increasing system size, the phase transition appears to be discontinuous. It is worth emphasizing that noise enhances the disintegration of spatial patterns, and thus the critical value of the competition rate $p_{c}$ increases with increasing system size; for example, $p_{c}=(2.3 \pm 0.1)$ for $L=512$ while $p_{c}=(3.1 \pm 0.1)$ for $L=700$. Figure~\ref{fragments}(a) shows single spirals for different system sizes at $E=2.0 \times 10^{-3}$ [which is below the critical $E_{c}$ value that can be inferred from Fig.~\ref{transitionheter}(b)], while Fig.~\ref{fragments}(b) shows pattern formation for different system sizes at $p=2.5$. Because competition effects at $L=200$ and $L=400$ in Fig.~\ref{transitionheter}(b) yield $E$ larger than $E_{c}$, multiple spirals are observed, i.e., the large spiral disintegrates to several smaller and less ordered spirals. Competition effects at $L=800$ and $L=1000$ in Fig.~\ref{transitionheter}(b), on the other hand, are below $E_{c}$, and therefore single-armed spirals are stable. Comparing the patterns at a given system size in Figs.~\ref{fragments}(a) and (b), one can also find that the wavelength of spirals decreases with increasing of $p$. To get a better insight into the fragmentation of single-arm spirals, without loss of generality, we focus on the system size $L=512$ where $p_{c}$ is $(2.3\pm 0.3)$. In Fig.~\ref{fragments}(c), we first draw typical spatial patterns as obtained over time for $p=2.5$ and $M=5.0\times 10^{-5}$. Starting from heterogeneous initial conditions, a single spiral emerges at $t=100000$. However, one species in the arm of the single spiral breaks out at $t=110900$ (marked by arrow), and shortly thereafter, two species break out at $t=111000$. When the system reaches the stationary state at $t=200000$, the single spiral is virtually completely fragmented and there is hardly any evidence left of its earlier existence. As shown in the top panels of Fig.~\ref{cri_predation}, while globally ordered spiral waves emerge for $p=0.1$ and $1.0$, the latter disintegrate for $p=10.0$. Further insights can be obtained by examining the spatial wavelength of spiral waves, defined as $\lambda=X/L$, where $X$ denotes the spatial distance between neighboring wave fronts with the same species. In particular, we find that the wavelength of spiral waves decreases as $p$ increases, and simultaneously the edges of the spirals separating different species become increasingly rough. We argue that these two facts eventually lead to the disintegration of globally ordered spiral waves for large enough values of $p$. Indeed, globally ordered spiral waves are no longer attainable for $p>p_{c}$. This assertion is based on extensive Monte Carlo simulations revealing the presence of spatially periodic structures. This phenomenon is also reproduced by the PDE model [see Eq.~(4)], as shown in the three bottom panels of Fig.~\ref{cri_predation}.

\begin{figure}
\begin{center} \includegraphics[width = 8.5cm]{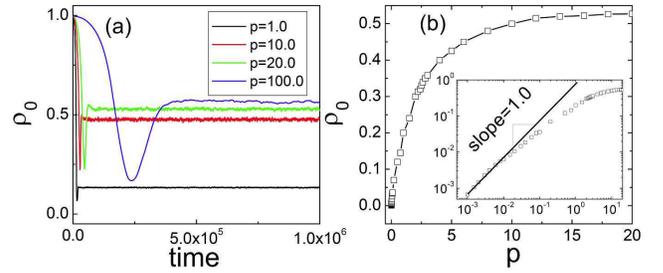}
\caption{\label{vacations}(color online) Panel (a) shows the density  of vacant sites $\rho_{0}$ evolving with time for different competition rates $p$. Panel (b) shows the stationary values of $\rho_{0}$ as function of the competition rate $p$, while the inset displays the data using log-log scale. Here $M=5.0\times 10^{-5}$, $N=512^2$ and $q=1.0$.}
\end{center}
\end{figure}

It is important to note that the competition rate $p$ affects not only the density of the three species responsible for the emergence of spatial patterns, but also the density of vacant sites, as shown in Fig.~\ref{vacations}(a). With increasing values of $p$, the density of vacant sites increases as well, but the effect is of saturating nature, as can be inferred from Fig.~\ref{vacations}(b). For small values of $p$ (at $p=10^{-2}$ for example), the reproduction process (note that here $q=1.0$) happens much faster than the competition process. There $\rho_{0}$ grows linearly with $p$ on a log-log scale (see inset of Fig.~\ref{vacations} (b)). However, for larger $p$ the reproduction process starts limiting the dynamics of $\rho_{0}$. Similarly as argued for results presented in Fig.~\ref{random}(d), $\rho_{0}$ therefore starts disobeying the linear relation with $p$, and one can imagine that $\rho_{0}$ will tend to $1.0$ as $p \rightarrow \infty$ because the system will then stay in the prepared initial conditions. However, for very large values of $p$ the density of vacant sites $\rho_{0}$ rises slowly indeed. In this case, the dynamics is almost entirely limited by reproduction, and patterns consisting of solitary wave fronts can be observed \cite{varea_jmb07}.

\begin{figure}
\begin{center} \includegraphics[width = 8.8cm]{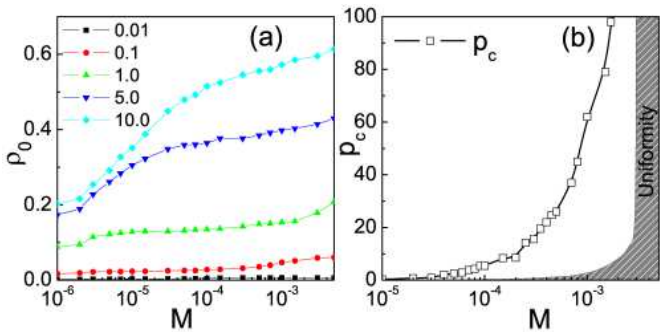}
\caption{\label{with_mobility}(color online) (a) The density of vacant sites $\rho_{0}$ as function of mobility $M$. (b) The critical competition rate $p_{c}$ as function of mobility $M$. The shadowed region denotes the emergence of an absorbing phase (uniformity), where two species go extinct. Here $N=512^2$ and $q=1.0$.}
\end{center}
\end{figure}

In Fig.~\ref{with_mobility}(a), we show further how the density of vacant sites varies with the critical competition rate $p_{c}$ as a function of mobility $M$. It can be observed that $\rho_{0}$ remains small for all $M$ at $p=0.01$ because the reproduction process happens much faster than the competition process. However, $\rho_{0}$ rises with increasing $M$ for large $p$ (at $p=10.0$ for example), as there the reproduction process happens much slower than the competition process, and hence the large mobility $M$ makes individuals prey more effectively. In Fig.~\ref{with_mobility}(b) we finally plot the full $p_c-M$ phase diagram, where the biodiversity region is depicted white and the absorbing single-species phase (uniformity) is depicted shadowed. In addition, the line in the white region delineates single spirals (left) and multiple spirals coexisting (right).

\section{Effects of reproduction rate on pattern formation}

Based on thus far presented results, we can conclude that the cyclic competition has positive effects on biodiversity, similarly to effects of reproduction rate reported first by Reichenbach \textit{et al.} \cite{reichenbach_n07}, while it has negative effects on pattern formation. It is therefore of interest to test briefly whether the reproduction rate might also have negative effects on spatial pattern formation. To address this, we here fix the competition rate to $p=1.0$, and perform Monte Carlo simulations of reactions (1)-(3) for different values of the reproduction rate $q$. As above, we start from the prepared initial state depicted schematically in Fig.~\ref{transitionheter}(a), since it promotes pattern formation, in particular the emergence of spirals.

\begin{figure}
\begin{center} \includegraphics[width = 8.5cm]{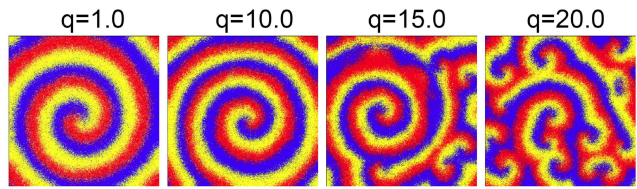}
\caption{\label{reproduction}(color online) Typical spatial patterns emerging for different reproduction rates $q$ at $M=5.0 \times 10^{-5}$. Prepared initial conditions, as depicted in Fig.~\ref{transitionheter}(a) were used. All patterns were observed after the system reached the stationary state. Here $N=512^2$ and $p=1.0$.}
\end{center}
\end{figure}

Figure~\ref{reproduction} presents typical spatial patterns emerging for different reproduction rates $q$ at $M=5.0 \times 10^{-5}$. At $q=1.0$, a globally stable spiral emerges, as we have already reported in the preceding section. With increasing values of of $q$, however, the wavelength of single arm spirals first decreases (compare $q=1.0$ and $q=10.0$), and for higher $q$ still (q=15.0 and 20.0) the single arm spirals break up and become more and more fragmented. This is very much in agrement with what we have reported above for the impact of $p$ by a fixed value of $q$, and it indeed confirms that increasing reproduction rates also negatively affect the emergence and stability of spatial patterns.

\section{Discussion}

Based on a biological rock-paper-scissors game, effects of different rates of cyclic competition on spatial pattern formation and biodiversity have been investigated. For low mobility of individuals where the three species coexist, we have examined pattern formation affected by cyclic competition rates between species, discovering that self-organized structures emerge if using prepared initial conditions. We have shown that globally ordered spiral waves, as were observed recently also in excitable systems \cite{perc_pre05, he_njp07}, can emerge from special heterogeneous initial conditions only if the competition effect is smaller than a critical threshold $E_{c}$. Since the introduced competition effect $E=p/L$ depends on the system size, the critical competition rate $p_{c}$ is therefore itself system size dependent. This is because the system size determines the level of noise in the system, which in turn may facilitate the disintegration of spatial patterns. We found that when approaching $p_{c}$ at a given system size, the borders separating the species forming the spirals become more and more rough (non-smooth). This is because the density of vacant sites increases, but also due to noise that is inherent for finite system sizes. Accompanying this is also a decrease of spatial wavelength of spiral waves $\lambda$, although the latter phenomenon bears no relevance for the impending disintegration of the spirals. At $p=p_{c}$ the disintegration of the globally ordered spirals occurs, resulting in predominantly disordered spatial portraits consisting of small fragmented spiral-like patterns. For random initial condition, on the other hand, our results are in agrement with those reported in Ref.~\cite{reichenbach_n07}, where biodiversity affected by reproduction rate has been studied first. We have also investigated phase transitions \cite{szolnoki_pre97} related to the extinction process evoked by different competition rates, where we found that, similarly as by the reproduction rate, increasing cyclic competition rates might have a positive effect on biodiversity due to the fragmentation of spirals, which prevents the patterns from outgrowing the system size.

It is worth noting that spatiotemporal patterns have been investigated extensively in the past in many different systems, ranging from chemical reactions on catalytic surfaces to propagating signals in aggregating microorganisms \cite{Mikhailov_pr06}. It has been shown that patterns in excitable systems emerge primarily due to the instabilities induced by the interplay between the fast excitatory and slow recovery variables \cite{perc_pre06}. This kind of mechanism explains well the spiral waves emerging in the Belousov-Zhabotinsky (BZ) reaction \cite{Vanag_JPCA00} and aggregating amoeba \emph{D. discoideum} \cite{Dormann_JCS97}, for example. Spirals in our systems, however, emerge because of the cyclic interaction between the three species, rather than differences in their dynamics. In addition, it is well known that the propagating signals and propagable interactions can lead to complex spatiotemporal patterns in systems describing Ca$^{2+}$ signaling in thalamocortical neurons \cite{Errington_JN2010} or interactions between predation and transport processes in a benthic nutrient-microorganism system \cite{Baurmann_MBE2004}. Similar mechanisms can also result in ordered spatiotemporal patterns occurring in the brain \cite{Benucci_N2007, Sharon_S2002,Pfurtscheller_CN2003} and heart \cite{Skanes_C1998}. These mechanisms, however, are significantly different from the cyclic competition presented in this paper, and the observed patterns and waves are accordingly dissimilar too. For example, sequential waves were observed for brain \cite{Benucci_N2007} and heart \cite{Skanes_C1998} tissue, yet single-armed spirals as we report presently are rarely reported. Altogether, our findings thus indicate that the competition rate in models of cyclically competing species is an important factor determining spatial pattern formation as well as mechanisms that are able to sustain it.

\section*{Acknowledgments}

This work was supported by the National Natural Science Foundation of China (Grant No. 11047012). Matja{\v z} Perc additionally acknowledges support from the Slovenian Research Agency (Grant No. Z1-2032).

%%\bibliography{egt}

\end{document}